\documentclass[12pt]{article}
\usepackage{amssymb,amsmath}

\title{A novel approach for solving the three-dimensional sine-Gordon equation}

\author{Sergey G Artyshev}

\date{Department of Applied Mathematics,\\
National Research Nuclear University MEPhI\\
31 Kashirskoe Shosse, 115409, Moscow, \\ Russian Federation \\
E-mail: \, SGArtyshev@mephi.ru}

\begin{document}

\maketitle

\begin{abstract}

A new way for finding analytical solutions of the three-dimensional sine-Gordon equation is presented. The method is based on the established relation between the solutions of the three-dimensional wave equation and solutions of the three-dimensional sine-Gordon equation. Some examples of the solutions thus obtained are show.
\end{abstract}

The sine-Gordon equation has been used to describe with a good approximation a number of physical phenomena \cite{Skyrme_58, Josephson_74, Kudryashov_08, Shohet_04}. The applications can additionally be extended under the condition: dimension of the equation greater than 1. However the methods already known for solving of the one-dimensional sine-Gordon equation cannot be used for finding a solutions of the n-dimensional sine-Gordon equation $( n>1 )$.

First of all, we show how a solution of the wave equation

\begin{equation}\label{a1}
\frac{\partial^2 F}{\partial t^2} - \frac{\partial^2 F}{\partial x^2} - \frac{\partial^2 F}{\partial y^2} - \frac{\partial^2 F}{\partial z^2} = 0 \qquad F=F \left( x, y, z, t \right)
\end{equation}
can be used for finding a solution of the three-dimensional sine-Gordon equation

\begin{equation}\label{a2}
\frac{\partial^2 u}{\partial t^2} - \frac{\partial^2 u}{\partial x^2} - \frac{\partial^2 u}{\partial y^2} - \frac{\partial^2 u}{\partial z^2} + \sin u = 0 \qquad u=u \left( x, y, z, t \right).
\end{equation}
The substitution \cite{Ouroushev_91}

\begin{equation}\label{a3}
u=4\tan^{-1}\sigma
\qquad \sigma=\sigma \left( x, y, z, t \right)
\end{equation}
leads to the following nonlinear partial differential equation for $ \sigma:$
\begin{equation}\label{a4}
(1+\sigma^{2})[\frac{\partial^2 \sigma}{\partial t^2} - \frac{\partial^2 \sigma}{\partial x^2} - \frac{\partial^2 \sigma}{\partial y^2} - \frac{\partial^2 \sigma}{\partial z^2}]-2\sigma[(\frac{\partial \sigma}{\partial t})^2 -(\frac{\partial \sigma}{\partial x})^2-(\frac{\partial \sigma}{\partial y})^2-(\frac{\partial \sigma}{\partial z})^2]=\sigma^3-\sigma.
\end{equation}
One possibility \cite{Ouroushev_91} for splitting equation \eqref{a4} into two is the following:

\begin{equation}\label{a5}
\frac{\partial^2 \sigma}{\partial t^2} - \frac{\partial^2 \sigma}{\partial x^2} - \frac{\partial^2 \sigma}{\partial y^2} - \frac{\partial^2 \sigma}{\partial z^2}=-\sigma
\end{equation}

\begin{equation}\label{a6}
(\frac{\partial \sigma}{\partial t})^2-(\frac{\partial \sigma}{\partial x})^2-(\frac{\partial \sigma}{\partial y})^2-(\frac{\partial \sigma}{\partial z})^2=-\sigma^2.
\end{equation}

\textbf{Proposition}
\textit{Let $F$ satisfies the wave equation \eqref{a1} and an equation
\begin{equation}\label{a7}
(\frac{\partial F}{\partial t})^2-(\frac{\partial F}{\partial x})^2-(\frac{\partial F}{\partial y})^2-(\frac{\partial F}{\partial z})^2=-1;
\end{equation}
then
\begin{equation}\label{a8}
\sigma=e^{F}
\end{equation}
is a solution of the system \eqref{a5}, \eqref{a6}.}

\textbf{Proof.} Combining
$$
(\frac{\partial F}{\partial t})^2-(\frac{\partial F}{\partial x})^2-(\frac{\partial F}{\partial y})^2-(\frac{\partial F}{\partial z})^2=[(\frac{\partial \sigma}{\partial t})^2-(\frac{\partial \sigma}{\partial x})^2-(\frac{\partial \sigma}{\partial y})^2-(\frac{\partial \sigma}{\partial z})^2]/\sigma^2
$$
and
$$
\frac{\partial^2 F}{\partial t^2} - \frac{\partial^2 F}{\partial x^2} - \frac{\partial^2 F}{\partial y^2} - \frac{\partial^2 F}{\partial z^2} =[\frac{\partial^2 \sigma}{\partial t^2} - \frac{\partial^2 \sigma}{\partial x^2} - \frac{\partial^2 \sigma}{\partial y^2} - \frac{\partial^2 \sigma}{\partial z^2}]/\sigma - (\frac{\partial F}{\partial t})^2+(\frac{\partial F}{\partial x})^2+(\frac{\partial F}{\partial y})^2+(\frac{\partial F}{\partial z})^2
$$
we get the proposition.

Consequently, if $ F \left( x, y, z, t \right) $  is a solution of system \eqref{a1},\eqref{a7}; then the function $ u \left( x, y, z, t \right) $ is a solution of \eqref{a2}. This are illustrated by following examples.

\textit{Example 1.} Let
$$
F\left(x,y,z,t\right)=t\sinh\psi+\left(x\cos\alpha+y\cos\beta+z\cos\gamma\right)\cosh\psi+C,
$$
where $(\cos\alpha,\cos\beta,\cos\gamma)$ is unit vector, and $\psi$ and $C$ are constants. Taking \eqref{a8} into account, solution \eqref{a3} becomes
$$
u\left(x,y,z,t\right)=4\tan^{-1}[e^{t\sinh\psi+(x\cos\alpha+y\cos\beta+z\cos\gamma)\cosh\psi+C}].
$$
We note that this formula denote as three-dimensional version well-know topological soliton of one-dimensional sine-Gordon equation\cite{Skyrme_58, Josephson_74, Kudryashov_08, Shohet_04}.

\textit{Example 2.} Let
$$
F\left(x,y,z,t\right)=h(x \pm t) + y\cos\alpha + z\sin\alpha,
$$
where $h$ is an arbitrary smooth function and $\alpha$ is constant; then \eqref{a3} is equivalent to
$$
 u\left(x,y,z,t\right)=4\tan^{-1}[e^{h(x \pm t ) + y\cos\alpha + z\sin\alpha}].
$$
We assume  that  $h=\ln(f)$ and $\alpha=0$ or $\alpha=\pi$, then
$$
u\left(x,y,t\right)=4\tan^{-1}[f(x \pm t)e^{\pm y}]
$$
is a solution of the two-dimensional sine-Gordon equation (see\cite{Ouroushev_91}).

\textit{Example 3.} The previous example may be generalized. Suppose a function $F\left(x,y,z,t\right)$ has the look:
$$
 F\left(x,y,z,t\right)=h(x\cos\alpha_1 + y\cos\beta_1 + z\cos\gamma_1 \pm t) + x\cos\alpha_2 + y\cos\beta_2 + z\cos\gamma_2,
$$
where  $(\cos\alpha_1, \cos\beta_1, \cos\gamma_1)$ and $(\cos\alpha_2, \cos\beta_2, \cos\gamma_2)$ are mutually orthgonal unit vectors, and $h$ is an arbitrary smooth function. Using \eqref{a3} we get new solution
$$
u\left(x,y,z,t\right)=4\tan^{-1}[e^{h(x\cos\alpha_1 + y\cos\beta_1 + z\cos\gamma_1 \pm t) + x\cos\alpha_2 + y\cos\beta_2 + z\cos\gamma_2}].
$$
Similarly, if $h=\ln(f)$, one obtains the solution
$$
u\left(x,y,z,t\right)=4\tan^{-1}[f(x\cos\alpha_1 + y\cos\beta_1 + z\cos\gamma_1 \pm t)e^{x\cos\alpha_2 + y\cos\beta_2 + z\cos\gamma_2}].
$$

\textit{Example 4.} Now suppose the function $ F\left(x,y,z,t\right) $ has the decomposition into the sum: $$ F\left(x,y,z,t\right)=\sum_{i=1}^k F_i(x,y,z,t). $$For example, when $ k=2, $  $$ F\left(x,y,z,t\right)=h_1(x\cos\alpha_1 + y\cos\beta_1 + z\cos\gamma_1 \pm t)
  + (x\cos\alpha_2 + y\cos\beta_2 + z\cos\gamma_2)/\sqrt { 2 } $$  $$ + h_2(x\cos\alpha_1 + y\cos\beta_1 + z\cos\gamma_1 \pm t) + (x\cos\alpha_3 + y\cos\beta_3 + z\cos\gamma_3)/ \sqrt { 2 },  $$where $ h_1, h_2 $ are arbitrary smooth functions; $(\cos\alpha_1, \cos\beta_1, \cos\gamma_1)$, $(\cos\alpha_2, \cos\beta_2, \cos\gamma_2) $ and $ (\cos\alpha_3, \cos\beta_3, \cos\gamma_3) $ are mutually orthgonal unit vectors. Substituting this expression into \eqref{a8} and the result into \eqref{a3}, we get new solutions of the three-dimensional sine-Gordon equation.

  The relations \eqref{a1} and \eqref{a7} can easily be checked by a direct calculation.

\end{document}